\documentstyle[editedvolume,psfig]{jd} 

\def\apj{ApJ }
\def\apjs{ApJS }

\def\aa{A\&A }

\def\mnras{MNRAS }

\def\etal{{\it et al.}}
\def\hii{H~{\sc ii}}

\def\heii{He~{\sc ii}}
\def\hbeta{\ifmmode {\rm H{\beta}} \else $\rm H{\beta}$\fi}
\def\msun{\ifmmode M_{\odot} \else $M_{\odot}$\fi}

\def\ga{\mathrel{\mathchoice {\vcenter{\offinterlineskip\halign{\hfil
$\displaystyle##$\hfil\cr>\cr\noalign{\vskip1.5pt}\sim\cr}}}
{\vcenter{\offinterlineskip\halign{\hfil$\textstyle##$\hfil\cr>\cr
\noalign{\vskip1.0pt}\sim\cr}}}
{\vcenter{\offinterlineskip\halign{\hfil$\scriptstyle##$\hfil\cr>\cr
\noalign{\vskip0.5pt}\sim\cr}}}
{\vcenter{\offinterlineskip\halign{\hfil$\scriptscriptstyle##$\hfil
\cr>\cr\noalign{\vskip0.5pt}\sim\cr}}}}}

\begin{opening}
\title{New light on the origin of nebular \heii\ emission 
in young starbursts}

\author{Daniel Schaerer}
\institute{Space Telescope Science Institute, 3700 San Martin Drive \\ 
	Baltimore, MD 21218, USA}

\end{opening}

\runningtitle{Origin of nebular \heii}

\begin{document}
 \vspace*{-1cm}

\section{Observations of nebular \heii\ emission} 
\vspace*{-0.2cm}
In the Local Group few \hii\ regions exhibit nebular 
\heii\ $\lambda$4686 emission, indicative of unusually high
excitation. Approximately eight such nebulae are
known, all located in low metallicity environments
(IC 1613, SMC, LMC) except the Galactic ring nebula 
G2.4+1.4 (Esteban \etal\ 1992).
The best studied case is the nebula S 3 surrounding the WO3
star DR 1 (Garnett \etal\ 1991, Kingsburgh \& Barlow 1995).
Except for two cases (N 44C, N159 F; Stasi\'nska \etal\ 1986,
Pakull \& Angebault 1986) the nebulae are asociated
with early WN (cf.\ Niemela \etal\ 1991) and WO stars.

So far approximately 50 to 60 extragalactic \hii\ regions
showing nebular \heii\ $\lambda$4686 are known (Campbell \etal\ 1986,
Izotov \etal\ 1994, 1997a). Most of them are found in \hii\ galaxies,
BCDs and related objects from low metallicity samples used for determinations 
of the primordial helium abundance. The most prominent case 
is the low metallicity (Z) record holder I Zw 18.
\heii\ emission is typically on the order of 1-5 \% of \hbeta.

\vspace*{-0.3cm}
\section{The origin of nebular \heii}
\vspace*{-0.2cm}
Different ionization mechanisms (photoionization by hot stars,
shock excitation, photoionization by X-rays) have been put 
forward to explain the required hard spectrum (Garnett \etal\
1991). 

The first quantitative approach to address this problem are
the evolutionary synthesis models of Schaerer (1996) and Schaerer
\& Vacca (1997), which include well tested evolutionary tracks 
and recent spherically expanding non-LTE atmosphere models 
appropriate for WR and O stars.{\em  These models predict 
nebular \heii\ emission due to hot WN and WC/WO stars} 
(\heii\ $\lambda$4686/\hbeta\ up to 0.01-0.02).
%
The classification of the ionizing sources agrees well with
the majority of the nebulae discussed above.
As shown by Schaerer (1996) the models are also in good agreement
with observations of WR galaxies showing the characteristic
broad WR emission (mostly \heii\ $\lambda$4686 and/or C~{\sc iv}
$\lambda$5808) and/or the presence of nebular \heii.

If we assume instantaneous bursts, no contamination from 
an underlying population and ionization bounded regions, 
the \hbeta\ equivalent width indicates the age of the population.
Comparing the observations with our recent synthesis models
(Schaerer 1996, Schaerer \& Vacca 1997)
we find that the vast majority of the extragalactic \hii\ 
regions with nebular \heii\ falls in the age range where 
WR stars are predicted to be present.
{\em This suggests that nebular emission may be related to WR
stars in all cases.}

\vspace*{-0.3cm}
\section{I Zw 18 -- or: new probes for massive star evolution at 
low Z}
\vspace*{-0.2cm}
The above suggestion is supported by the recent discovery of 
WC and WN stars in I Zw 18 (Izotov \etal, 1997b, Legrand \etal\ 1997).
Furthermore the observed nebular \heii\ emission can be fully explained
if the ionizing flux of DR 1 (Kingsburgh \& Barlow) is representative
for the observed WC stars in I Zw 18.

\begin{figure*}[htb] 
\centerline{
\psfig{width=6.5cm,figure=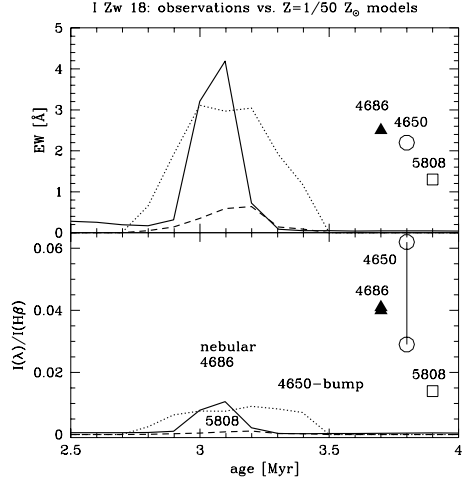}
}
\caption{ 
Predicted nebular \heii\ $\lambda$4686 (solid), 4650 bump (dotted),
5808 bump (dashed) for an instantaneous burst with a Salpeter IMF and
$M_{\rm up}=150 \, \msun$ (top: equivalent widths, bottom: line intensity
$I/\hbeta$).
Observations from Izotov et al.\ (1997b) and Legrand et al.\ (1997)
are shown on the right side. As discussed in the text $I($4650$)/\hbeta$ 
shows important differences between the two observations.
}
\end{figure*}

To examine this case in more depth we have calculated new evolutionary 
tracks and synthesis models for Z=0.0004 of I Zw 18. 
We obtain a mass limit for the formation of WR stars of $M_{\rm WR}
\ga 90 \, \msun$.
Predictions for the WR bumps and nebular \heii\ $\lambda$4686
for an instantaneous burst with a Salpeter IMF extending to 150 \msun\
are shown in Figure 1.
The line intensities (bottom) seem to indicate a difficulty for the models.
However, the observed equivalent widths are in fair agreement, which
shows that the above comparison is hampered by different spatial 
extensions of the gas and stars, as also recognized by Izotov et al.
In particular the nebular 4686 can well be explained with a Salpeter IMF
extending to large a, but not unreasonable, upper mass cut-off.
\newpage
{\em Although not being a conclusive proof, the results obtained so far
support the hypothesis that WR stars are responsible for
nebular 4686 emission.}

The observations of Izotov et al.\ (1997b) and Legrand et al.\ differ
in the region of the 4650-4686 WR bump. While the former find broad
emission centered at 4686 \AA, the latter exclude any broad 4686 emission
and detect a bump at 4645 \AA, attributed to C~{\sc iii} emission from
WC stars. These differences could e.g.\ be due to the different slit 
positions. As a consequence the exact content of WN and WC stars
in I Zw 18 remains uncertain. 
 
The observed strength of the total 4650-4686 WR bump and C~{\sc iv}
$\lambda$5808 are well reproduced by the models.
However, we predict broad O~{\sc v} $\lambda$5590 emission with an 
equivalent width of approximately 1 \AA, which does not seem to be 
observed.
Indeed at this low Z the single star evolutionary models predict WN, 
no WC, but instead oxygen rich stars (WO).
This is due to the fact that at low Z the He-burning core (identified
with WC or WO stars according to the C, O, and He composition)
is revealed at a more chemically advanced stage given the 
low expected mass loss.
Roche lobe overflow in binary systems may have lead to the formation 
of WC stars. Alternatively the presence of WC stars in I Zw 18 might indicate
a deficiency in the evolutionary tracks, which have so far not 
been tested at metallicities below the SMC. 
Systematic studies of the WR (WN, WC, and possibly also WO !) and O 
star populations in very low metallicity \hii\ galaxies are very promising
in this respect.
Future analysis of stellar populations
in such environments should greatly improve our knowledge of massive 
star evolution in the early universe.

\acknowledgements
{\small
Last minute support from the IAU and a travel grant
from the Swiss Society for Astronomy and Astrophysics are greatly 
acknowledged.
}


\end{document}